# Tracking yeast pheromone receptor Ste2 endocytosis using fluorogen-activating protein tagging

Anita Emmerstorfer-Augustin[a], Christoph M. Augustin[b], Shadi Shams[a], and Jeremy Thorner[a,*]
[a]Division of Biochemistry, Biophysics and Structural Biology, Department of Molecular and Cell Biology, and [b]Department of Mechanical Engineering, University of California, Berkeley, Berkeley, CA 94720-3202

**ABSTRACT** To observe internalization of the yeast pheromone receptor Ste2 by fluorescence microscopy in live cells in real time, we visualized only those molecules present at the cell surface at the time of agonist engagement (rather than the total cellular pool) by tagging this receptor at its N-terminus with an exocellular fluorogen-activating protein (FAP). A FAP is a single-chain antibody engineered to bind tightly a nonfluorescent, cell-impermeable dye (fluorogen), thereby generating a fluorescent complex. The utility of FAP tagging to study trafficking of integral membrane proteins in yeast, which possesses a cell wall, had not been examined previously. A diverse set of signal peptides and propeptide sequences were explored to maximize expression. Maintenance of the optimal FAP-Ste2 chimera intact required deletion of two, paralogous, glycosylphosphatidylinositol (GPI)-anchored extracellular aspartyl proteases (Yps1 and Mkc7). FAP-Ste2 exhibited a much brighter and distinct plasma membrane signal than Ste2-GFP or Ste2-mCherry yet behaved quite similarly. Using FAP-Ste2, new information was obtained about the mechanism of its internalization, including novel insights about the roles of the cargo-selective endocytic adaptors Ldb19/Art1, Rod1/Art4, and Rog3/Art7.

## INTRODUCTION

G protein–coupled receptors (GPCRs) are the most numerous and diverse superfamily of cell-surface receptors (Davenport *et al.*, 2013; Vass *et al.*, 2018). GPCRs share a common structural organization, with an extracellular N terminus, seven transmembrane-spanning domains, and a cytoplasmic C terminus (Preininger *et al.*, 2013; Lee *et al.*, 2015), and trigger downstream signal transduction using similar mechanisms (Lohse and Hofmann, 2015; Hilger *et al.*, 2018). The first genes isolated for GPCRs that respond to peptide agonists were the pheromone receptors, Ste2 and Ste3, of budding yeast *Saccharomyces cerevisiae* (Burkholder and Hartwell, 1985; Nakayama *et al.*, 1985; Hagen *et al.*, 1986). Since their identification, study of these receptors has provided numerous path-finding insights about GPCR-initiated signaling (Dohlman and Thorner, 2001; Naider and Becker, 2004; Konopka and Thorner, 2013). Ste2 resides in the plasma membrane (PM) of *MAT***a** cells and binds α-factor, the 13-residue pheromone secreted by *MAT*α cells, thereby initiating a cascade of events (reviewed in Merlini *et al.* [2013] and Alvaro and Thorner [2016]) that lead to activation of a mitogen/messenger-activated protein kinase whose actions result in cell-cycle arrest in the G1 phase, cause highly polarized growth (called "shmoo" formation) (Madden and Snyder, 1998), and induce the transcription of genes required to prepare a *MAT***a** haploid for cell and nuclear fusion with a *MAT*α haploid.

However, should a *MAT***a** cell fail to conjugate with a *MAT*α partner, among the pheromone-induced gene products are factors that exert feedback mechanisms that limit the duration of signaling, promote recovery from pheromone-induced G1 arrest, and permit resumption of mitotic proliferation—a striking example of the survival value to this yeast species of what is referred to in evolutionary theory as "bet-hedging" (Grimbergen et al., 2015). Proteins upregulated by α-factor induction in *MATa* cells that act to dampen signaling at the receptor level include the following: Bar1, an α-factor-degrading protease; Sst2, an RGS protein that promotes nucleotide hydrolysis when GTP is bound to Gpa1 (the α subunit of the receptor-associated heterotrimeric G protein); and, Gpa1 itself but not its cognate Gβγ (Ste4-Ste18) complex, which, by mass action, allows for recapture of free Gβγ, that, in this system, is responsible for triggering signal initiation downstream of receptor activation (Merlini et al., 2013; Alvaro and Thorner, 2016).

Ste2 itself undergoes basal endocytosis and more rapid ligand-induced internalization (Jenness and Spatrick, 1986; Zanolari and Riezman, 1991). On α-factor binding, Ste2 becomes hyperphosphorylated on its cytoplasmic tail (Reneke et al., 1988), which promotes its ubiquitinylation (Hicke et al., 1998) by the PM-associated ubiquitin ligase (E3) Rsp5 (Dunn and Hicke, 2001), which installs K63-linked polyubiquitin chains (Belgareh-Touzé et al., 2008; Lauwers et al., 2009), and ubiquitinylated Ste2 then is recognized by the cargo receptors that mediate clathrin-dependent endocytosis (Shih et al., 2002; Toshima et al., 2009). The resulting Ste2-containing endosomes are directed to the multivesicular body (Odorizzi et al., 1998) and then to the vacuole where the receptor is degraded (Schandel and Jenness, 1994; Gabriely et al., 2007). However, Rsp5 is unable to associate directly with the integral PM proteins that are its clients; cargo-selective adaptor proteins, the α-arrestins, serve as molecular matchmakers to tether Rsp5 to its targets (Lin et al., 2008; Nikko and Pelham, 2009; Becuwe et al., 2012). In the case of Ste2, we have shown that 3 of the 14 known α-arrestins in yeast, Ldb19/Art1, Rod1/Art4, and Rog3/Art7, make the most major contributions to Ste2 down-regulation (Alvaro et al., 2014).

Certain of the above conclusions were reached using bound radioactive α-factor as an indirect proxy for its cognate receptor. More recently, functional versions of Ste2 tagged at its C terminus with GFP or other fluorescent protein have been used to monitor its localization. However, due its constitutive endocytosis, high background fluorescence accumulates in the vacuole, causing significant signal-to-noise problems in visualizing the population of Ste2 at the PM and other cellular locations (Alvaro et al., 2014; Ballon et al., 2006). One strategy to surmount fluorescence accumulation in the vacuole/lysosome has been to use so-called superecliptic pHluorin as the tag, which rapidly loses fluorescence when pH < 6 (Prosser et al., 2016). This tactic has worked well for Ste3 but not Ste2 (Prosser et al., 2015). Moreover, significant questions about Ste2 dynamics and intracellular trafficking remain to be addressed, especially after cells are exposed to α-factor. For example, although in naive cells (i.e., not treated with pheromone), Ste2 is delivered rather uniformly to the PM, very rapidly after pheromone addition, essentially all of the detectable α-factor binding sites disappear with a half-time of ∼7 min (Jenness and Spatrick, 1986; Reneke et al., 1988; Rohrer et al., 1993); yet, concomitant with this apparent loss, a prominent "cap" of receptor becomes concentrated at the tip of the shmoo projection (Ballon et al., 2006). Based on experiments in which actin-based secretion was presumably blocked by treatment with latrunculin A (LatA) or a *myo2-16*ts allele, it was reported that this polarization of the yeast pheromone receptor requires its internalization but not actin-dependent secretion (Suchkov et al., 2010). Various explanations were offered for this surprising conclusion, such as biased fusion of vesicles containing Ste2-GFP, tendency of Ste2 to form dimers, local changes in the PM composition that could attract or stabilize receptor clusters, or faster internalization of the receptor at locations in the cell other than at the shmoo tip (Suchkov et al., 2010). However, given that *STE2* is a pheromone-induced gene (Hartig et al., 1986) and that actin cables direct vesicle-mediated secretion of all other membrane cargo yet examined to the shmoo tip (Liu and Bretscher, 1992; Lillie and Brown, 1994; Garrenton et al., 2010), formation of this cap of receptors likely depends on actin-dependent secretion of newly made receptors, rather than solely on clustering of preexisting receptors at the shmoo tip.

To address such issues, it would be advantageous to follow only Ste2 molecules present at the cell surface at the time of agonist engagement. Also, labeling the Ste2 N terminus would obviate concerns that bulky C-terminal tags could interfere with negative regulators and endocytic effectors (Dohlman and Thorner, 2001; Wolfe and Trejo, 2007; Kim et al., 2012), which all act from the cytoplasm. A method to achieve these goals is to tag an integral membrane protein with an exocellular fluorogen-activating protein (FAP) (Szent-Gyorgyi et al., 2008; Holleran et al., 2010; Li et al., 2017). A FAP tag is a relatively small (∼200 residues), human single-chain antibody engineered to bind tightly a cell-impermeable dye (fluorogen), which thereby is converted from a nonfluorescent to a fluorescent state. FAP tagging has allowed visualization of receptor internalization in mammalian cells; but its use to follow endogenous PM proteins in yeast, which possesses a cell wall, had not been tested. As described here, we successfully generated functional FAP-tagged Ste2, established conditions that permit its stable expression, and were then able, for the first time, to monitor both basal and ligand-induced receptor internalization of only those molecules at the cell surface and thereby gain new insights about the routes of endocytic trafficking taken by this receptor, as well as to reveal distinct roles for the α-arrestins Ldb19, Rod1, and Rog3.

## RESULTS

### Construction and validation of FAP-tagged Ste2

Two FAP tags—FAPα2 (binds cell-impermeable malachite green derivatives and emits red fluorescence) and FAPβ1 (binds cell impermeable thiazole orange derivatives and emits green fluorescence)—were developed initially (Supplemental Figure S1A), wherein the N terminus is marked with an influenza virus hemagglutinin (HA) epitope and the C terminus with a Myc epitope, and in both of which the signal peptide of the *kappa* light chain (Igκ) of human immunoglobulin G (IgG) directs secretion (Szent-Gyorgyi et al., 2008). We fused each FAP cassette in-frame to the methionine start codon (ATG) of the *STE2* open reading frame (ORF) that was also tagged in-frame at its C terminus with an octapeptide epitope (DYKDDDDK) from the Gene 10 protein of *Escherichia coli* bacteriophage T7 (FLAG tag) and a $(His)_6$ tract, which, as we demonstrated previously, do not alter any measurable function of this receptor (David et al., 1997). We retained the entire Ste2 N-terminal sequence in these constructs because of existing evidence that this portion of the receptor is important for its surface expression and proper folding (Uddin et al., 2012, 2016, 2017). These chimeric constructs, expressed from the *STE2*$_{prom}$ on a *CEN* plasmid, as well as a control expressing Ste2-FLAG-$(His)_6$ from the same vector, were introduced into *MATa ste2Δ* cells. Immunoblotting revealed that both FAP-containing proteins were expressed and, compared with the Ste2-FLAG-$(His)_6$ control (Supplemental Figure S1B, left), exhibited the increase in size expected for these chimeric receptors (Supplemental Figure S1B, right). Thus, the human FAP sequences were no impediment to transcription and translation in yeast. However, reproducibly, the FAPα2-Ste2 construct was expressed at a significantly higher level than FAPβ1-Ste2 (Supplemental Figure S1B, right). Moreover, when incubated briefly with structs was integrated into the *STE2* locus and expressed from the endogenous *STE2* promoter. The MFα1$^{(1-83)}$-Igκ-FAPα2-Ste2 construct (see Supplemental Table S2 for full nucleotide sequence), which contains most of the prepro-leader sequence in the precursor of the secreted pheromone α-factor (Fuller *et al.*, 1988), emerged as the candidate that yielded the best combination of robust expression (Supplemental Figure S2B), full retention of pheromone-responsive receptor signaling capacity (Supplemental Figure S2C), and maximal fluorescence on fluorogen binding (Supplemental Figure S2D). This construct (hereafter "FAP-Ste2") was used for all further analyses.

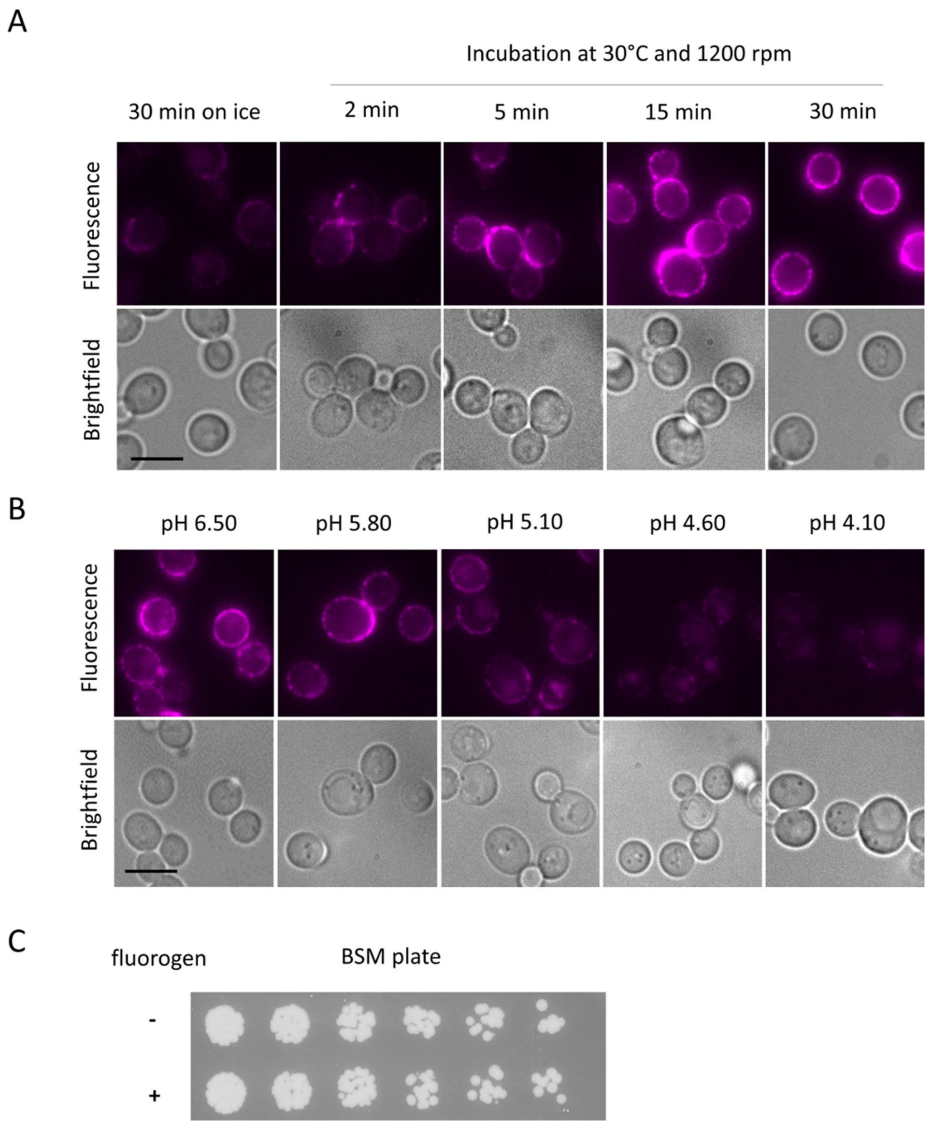

To establish the utility of FAP-Ste2 for monitoring receptor localization, we first optimized the conditions for its labeling. Unlike FAP-tagged proteins in animal cells, which generate a robust fluorescent signal when incubated with fluorogen on ice for 5 min (Holleran *et al.*, 2010, 2012; Boeck and Spencer, 2017), we found that maximal fluorogen binding to FAP-Ste2 required incubation for 15 min even at 30°C (Figure 1A), the optimal temperature for yeast cell growth, suggesting that the dye is slow to diffuse through the yeast cell wall. *Saccharomyces cerevisiae* prefers to grow at somewhat acidic pH. Whether cells were propagated at a given pH and then incubated with fluorogen at the same pH (Figure 1B), or pregrown at pH 6.5 and then shifted to medium at a different pH and then incubated with fluorogen (unpublished data), stable labeling was observed only at values approaching pH 6. Therefore, in all subsequent experiments, cells were grown in medium buffered at pH 6.5. Examination of viable titer after exposing FAP-Ste2-expressing cells to fluorogen at pH 6.5 for 15 min at 30°C demonstrated that exposure to the dye under these conditions had no toxic effect (Figure 1C).

**FIGURE 1:** Optimization of fluorogen binding to FAP-Ste2. (A) Cells (yAEA152) expressing FAP-Ste2 from the endogenous *STE2* locus were grown to mid–exponential phase in BSM, incubated with fluorogen (0.4 mM final concentration) either on ice without agitation or at 30°C with agitation (1200 rpm) for the time periods indicated, washed and collected by brief centrifugation, and viewed by fluorescence microscopy (top panels) and bright field microscopy (bottom panels), as described under *Materials and Methods*. Scale bar, 5 μm. (B) As in A, except the cells were propagated in BSM buffered at the indicated pH values (with either 100 mM phosphate or 50 mM succinate, as appropriate), incubated with fluorogen for 15 min at 30°C, and then imaged. (C) Portions of the same culture as in A were incubated for 15 min at 30°C in the absence (−) or presence (+) of fluorogen, and then samples of a set of fivefold serial dilutions were spotted using a multiprong inoculator on an agar plate containing BSM, and, after incubation for 48 h at 30°C, the resulting growth was recorded.

### Maintenance of intact FAP-Ste2

As another means to confirm that FAP-Ste2 retains receptor function, we used as a probe fluorescent Alexa Fluor 488–labeled α-factor (488-αF), prepared as described (Toshima *et al.*, 2006). When incubated with cells lacking Ste2 (Figure 2A, left), no significant binding was detectable, whereas for control cells expressing Ste2-FLAG-(His)$_6$ (Figure 2A, middle), prominent decoration of the cell surface was observed. Likewise, for cells expressing FAP-Ste2 (Figure 2A, right), prominent decoration of the cell surface was observed, which, reassuringly, was largely congruent with the FAP signal. As expected, on further incubation, the 488-αF initially bound to both the Ste2-FLAG-(His)$_6$-expressing cells (Figure 2A, middle) and the FAP-Ste2-expressing cells (Figure 2A, right) was trafficked to endocytic compartments and then apparently degraded. However, the majority of the FAP signal was not simultaneously internalized (Figure 2A, right). Because their cognate fluorogens, only the cells expressing the FAPα2-Ste2 construct yielded a readily detectable fluorescent signal and that fluorescence was located, as expected, largely at the cell periphery (Supplemental Figure S1C).

To determine whether we could improve surface expression of FAPα2-Ste2 while retaining the proper folding and function of both its FAP and receptor domains, the secretory signal sequences of three endogenous yeast proteins (MFα1, Ste2, and Suc2) were installed, either in place of or immediately upstream of the Igκ signal peptide (Supplemental Figure S2A), as described in detail in the Supplemental Material. Each of these different signal peptide con-

**2722** | A. Emmerstorfer-Augustin *et al.* Molecular Biology of the Cell

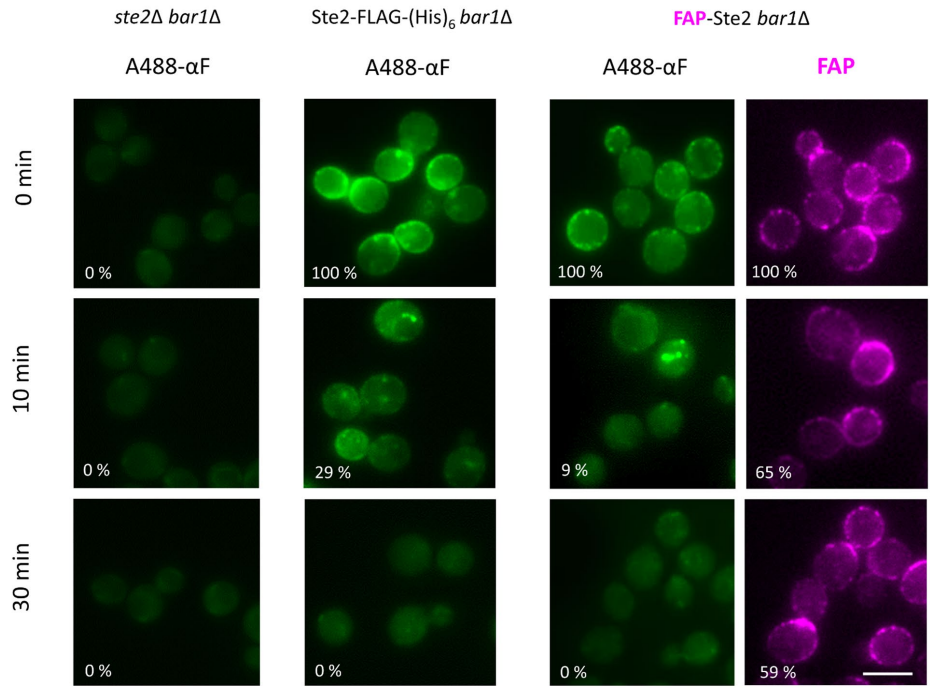

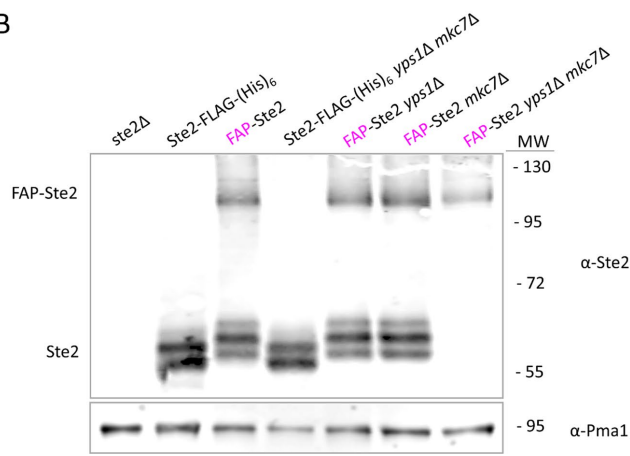

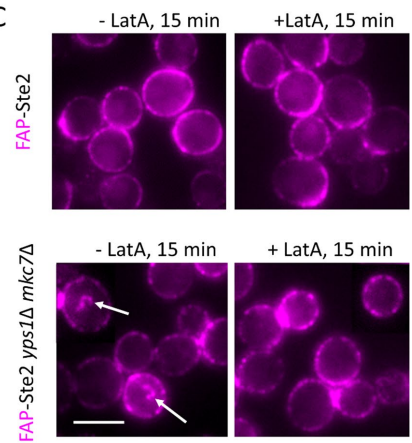

FIGURE 2: Absence of yapsins preserves full-length endocytosis-competent FAP-Ste2. (A) Strain DK102 (*ste2Δ bar1Δ*) or otherwise isogenic derivatives expressing from the endogenous *STE2*$_{prom}$, either Ste2-FLAG-(His)$_6$ (yAEA265) or FAP-Ste2 (yAEA261), were incubated with A488-αF on ice for 1.5 h in medium lacking glucose and then washed and shifted to glucose-containing medium at 30°C, and samples were removed at the indicated times and viewed by fluorescence microscopy. The cells expressing FAP-Ste2 were prelabeled with fluorogen under standard conditions (0.4 mM dye; 15 min, 30°C, pH 6.5) prior to incubation with A488-αF. Value (%) in the bottom left corner of each image represents the average pixel intensity (*n* ≥ 200 cells per sample) of A488-αF or FAP-Ste2 at the cell periphery, relative to the starting intensity for each strain, quantified using CellProfiler, as described under *Materials and Methods*. Scale bar, 5 μm. (B) Strain JTY4470 (*ste2Δ*) and otherwise isogenic *yps1Δ* or *mkc7Δ* single mutant derivatives or a *yps1Δ mkc7Δ* double mutant derivative (Table 1), expressing from the endogenous *STE2*$_{prom}$ either Ste2-FLAG-(His)$_6$ or FAP-Ste2, as indicated, were grown to early exponential phase at 20°C, harvested, and lysed, and membrane proteins were extracted, resolved by SDS–PAGE, and analyzed by immunoblotting with anti-Ste2 antibody, as described under *Materials and Methods*. Loading control, Pma1 detected on the same immunoblots using anti-Pma1 antibody. MW, marker proteins (kDa). (C) Samples of a *YPS1*⁺ *MKC7*⁺ strain (yAEA152) or an otherwise isogenic *yps1Δ mkc7Δ* strain (yAEA359), each expressing FAP-Ste2, were treated, as indicated, with either vehicle alone (ethanol) or LatA in ethanol (100 μM final concentration) and then exposed to fluorogen as in A and viewed by fluorescence microscopy. Arrows, internalized vesicles containing FAP-Ste2. Scale bar, 5 μm.

binding and internalization of 488-αF is strictly receptor dependent and fluorescence of the chimera receptor is strictly FAP dependent, our observations suggested that FAP-Ste2 was being severed by proteolysis between its two domains. Indeed, immunoblot analysis (Figure 2B; see also Supplemental Figure S2B) confirmed that the majority of the FAP-Ste2 was suffering such cleavage. Given that the junction between the FAP tag and the receptor lies in the periplasmic space between the PM and the cell wall, we suspected that members of a family of extracellular, glycosylphosphatidylinositol (GPI)-anchored aspartyl proteases, known as yapsins (Krysan et al., 2005; Gagnon-Arsenault et al., 2006), might be responsible for this proteolysis. Indeed, immunoblotting documented that FAP-Ste2 was completely stable in a strain in which the genes coding Yps1 and Mkc7, two major paralogous yapsins, were deleted (Figure 2B). Moreover, unlike in wild-type cells, in the yps1Δ mkc7Δ cells, even basal endocytosis of FAP-Ste2 was readily observable, which was, as expected, actin dependent because it was blocked by the presence of LatA (Figure 2C). Hence, in all subsequent experiments, we used yps1Δ mkc7Δ cells expressing FAP-Ste2.

### FAP-Ste2 visualization of the PM receptor pool is superior to Ste2-EGFP or Ste2-mCherry

Although 30°C is the optimal growth temperature for yeast, we noted that in the original protocol using yeast surface display to develop the FAP tags, the cells were always propagated at 20°C (Szent-Gyorgyi et al., 2008). Hence, we examined whether the folding, stability, and/or delivery of FAP-Ste2, even in yps1Δ mkc7Δ cells, might be further enhanced at the lower temperature. We found by three independent, but complementary, criteria—namely intensity of the fluorogen-generated signal (Figure 3, A and B), immunoblot analysis (Figure 3C), and bioassay of pheromone responsiveness (Figure 3D)—that growth at 20°C yielded an approximately twofold increase in FAP-Ste2 over that seen at 30°C. Moreover, remarkably, the same trends also were seen, in every case, for Ste2-FLAG-(His)$_6$, Ste2-GFP(F64L S65T) mutant (EGFP), and Ste2-mCherry (Figure 3).

Most satisfyingly, however, regardless of the temperature, the fluorescent signal at the PM observed with FAP-Ste2 is much more distinct than for Ste2-EGFP and markedly more clear than for Ste2-mCherry (Figure 3A). Moreover, on initial incubation with fluorogen, the signal from internal compartments is minimal for the cells expressing FAP-Ste2, whereas there is persistent and massive accumulation of background fluorescence in the vacuole in the cells expressing Ste2-EGFP and Ste2-mCherry (Figure 3A). As expected, because the cells expressed each of these constructs in the same way (integrated at the STE2 locus on chromosome VI), the degree of stochastic variation in relative signal brightness from cell to cell was quite similar for FAP-Ste2, Ste2-EGFP, and Ste2-mCherry (Figure 3B). Furthermore, we determined that FAP-Ste2 expressed in yps1Δ mkc7Δ cells has nearly the same affinity for α-factor as other Ste2 variants. For this purpose, we introduced an sst2Δ mutation, which makes cells more sensitive to α-factor and thus allows measurement of pheromone response by the halo bioassay over a broader and more linear range of α-factor concentrations (Reneke et al., 1988; Alvaro et al., 2014). Such dose–response curves showed that the half maximal inhibitory concentration (IC$_{50}$) for sst2Δ yps1Δ mkc7Δ cells expressing FAP-Ste2 was only approximately fourfold higher than for the same cells expressing Ste2-FLAG-(His)$_6$ (Supplemental Figure S3).

### Direct visualization of basal and ligand-induced receptor endocytosis

Having established optimal expression and labeling conditions, we were able to monitor, uniquely and for the first time, the dynamics of just the population of cell-surface Ste2 molecules that are exposed to the extracellular milieu. MATa yps1Δ mck7Δ cells expressing FAP-Ste2 were propagated at 20°C and, to block any endocytosis during incubation with fluorogen, the cells were treated with LatA. Synchronous initiation of receptor internalization in the absence and presence of α-factor was then initiated by washing out the LatA. The resulting fluorescent images were striking. In the absence of pheromone (Figure 4A, top panels), prominent PM fluorescence persisted in a significant fraction of the cells for at least 45 min and the appearance of substantial fluorescence in endosomes took ~30 min. In marked contrast, in the presence of pheromone (Figure 4A, bottom panels), significant fluorescence in endosomes was visible by 5 min and PM fluorescence was approaching undetectable within 10 min. To determine internalization rate, we used CellProfiler to measure the average pixel intensity of the fluorescence only at the cell periphery in cells (n = 150–200) at each time point. These data yielded a half-life for receptor removal from the PM via basal endocytosis of ~25 min, whereas in the presence α-factor the half-time for internalization was only ~6 min, indicating that the rate of receptor endocytosis was accelerated four- to fivefold by ligand binding. Our data are in good general agreement with the rates of constitutive and pheromone-induced Ste2 endocytosis determined in other ways (Jenness and Spatrick, 1986; Reneke et al., 1988; Zanolari and Riezman, 1991; Hicke et al., 1998; Toshima et al., 2006).

### Newly made receptors cap the tip of the mating projection

Having validated in the various ways documented above that FAP-Ste2 provided a reliable readout of authentic receptor behavior, we sought to use this tool to address some unresolved issues about Ste2. As observed originally using quantification of α-factor binding sites (Jenness and Spatrick, 1986), and as we have documented directly here (Figure 4), yeast cells exposed to pheromone rapidly internalize the receptor. However, by 30 min after initial exposure to pheromone, fresh α-factor binding sites appear and new protein synthesis is required for their appearance (Jenness and Spatrick, 1986), and, concomitantly, receptors accumulate at the shmoo tip, as visualized using Ste2-mCherry (Ballon et al., 2006) or Ste2-GFP (Arkowitz, 1999; Venkatapurapu et al., 2015) (Figure 5, left). Suchkov et al. (2010) reported that this marked Ste2 polarization requires its internalization, but not actin-dependent secretion, implying, among other potential explanations, that this distribution could arise from preferential endocytosis of the receptor except at the shmoo tip rather than from de novo synthesis and insertion of new receptors at the shmoo tip. However, our findings (Figure 4) already suggested that there was no region of the PM where FAP-Ste2 was "immune" to ligand-induced endocytosis. To address this question by an alternative approach, we exposed MATa yps1Δ mck7Δ cells expressing FAP-Ste2 to excess α-factor for 3 h to give sufficient time for the cells to form prominent shmoos and to ensure that all preexisting surface-exposed FAP-Ste2 would be long since internalized and completely destroyed (see Figure 4), then added LatA to block actin-based secretion or endocytosis, and, finally, incubated the cells with fluorogen. Exposure to fluorogen at this stage revealed prominent concentration of the FAP-Ste2 molecules made during the pheromone treatment at the shmoo tip (Figure 5, right), demonstrating unequivocally that these receptor "caps" arise from de novo synthesis and insertion of newly made receptor molecules at this location, in agreement with similar conclusions reached using less direct methods (Ayscough and Drubin, 1998; Moore et al., 2008).

### Arf-GAP Glo3 is required for trafficking of endocytosed Ste2 to the vacuole

Neither our approach for following surface Ste2 directly nor prior studies (Tan et al., 1993; Schandel and Jenness, 1994) provide any evidence that Ste2 is recycled from endosomes back to the PM as an alternative to its delivery to the vacuole either during its basal endocytosis (Supplemental Figure S4) or after agonist-induced internalization (Figure 4). Yet it has been reported recently (Kawada et al., 2015), on the basis of the rate of uptake of [$^{35}$S]α-factor, that yeast cells lacking the Arf-GAP Glo3 internalize Ste2 somewhat less efficiently than wild-type (WT) cells but have more prominent defects in the late endosome-to-*trans*-Golgi network transport pathway, and, therefore, Ste2 endocytosed in *glo3Δ* cells is sorted to the vacuole rather than recycled to the PM. Instead of tracking the receptor itself, the conclusions of Kawada *et al.* (2015) were reached mainly using α-factor covalently labeled with a bulky fluorescent dye on its sole Lys residue (K7), which their own prior work demonstrated reduces its affinity for Ste2 by at least 50-fold (Toshima *et al.*, 2006). Moreover, Kawada *et al.* (2015) also reported, using [$^{35}$S]α-factor labeled in its sole Met residue (M12), that, compared with wild-type cells, glo3Δ mutants exhibited, for unexplained reasons, a marked decrease in initial surface binding of pheromone, indicating a drastic reduction in the number of Ste2 molecules at the cell surface.

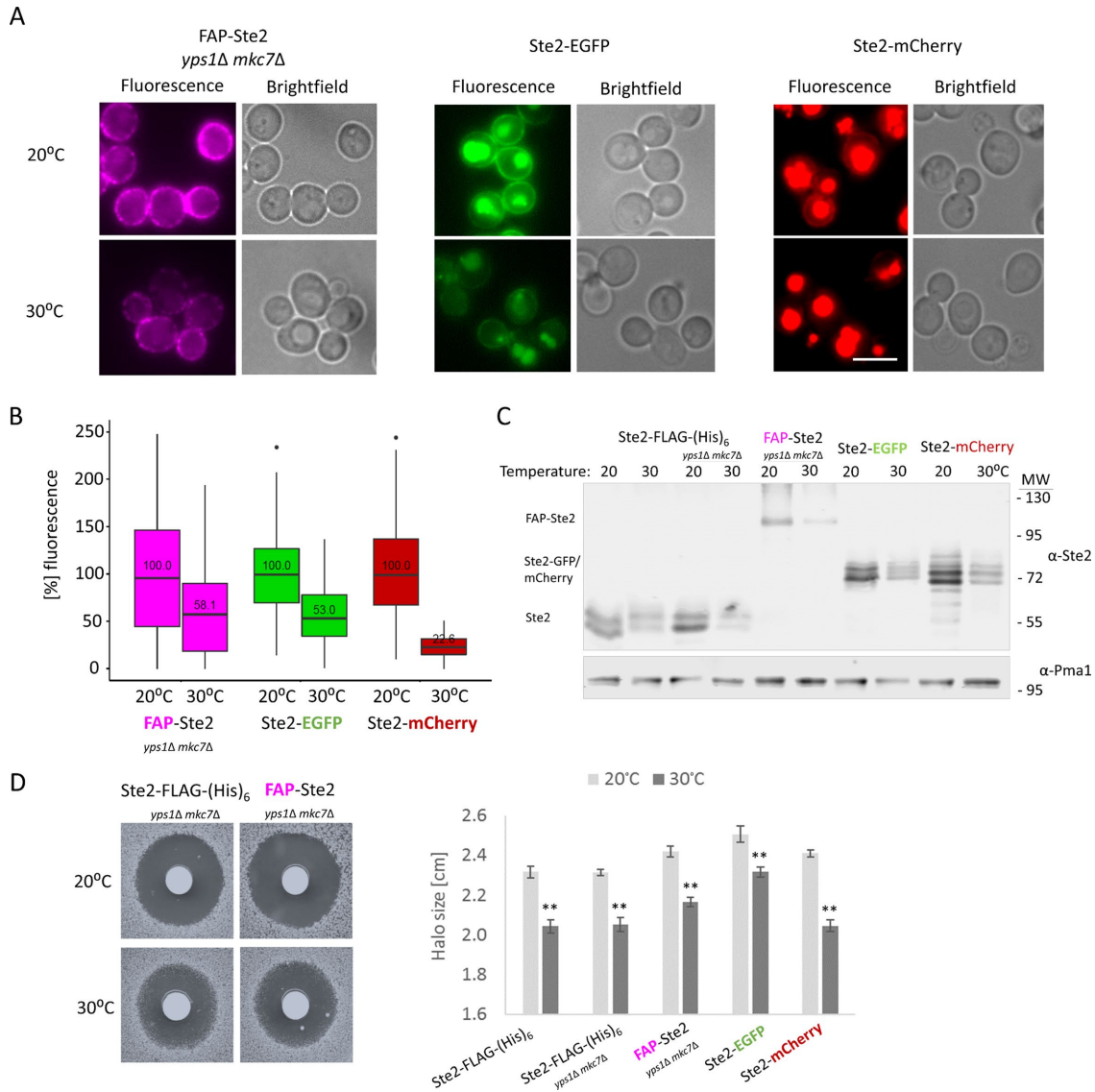

**FIGURE 3:** Comparison of FAP-Ste2 to Ste2-EGFP and Ste2-mCherry at two different temperatures. (A) A *MATa yps1Δ mkc7Δ* strain (yAEA359) expressing FAP-Ste2 from the *STE2* locus, and a *MATa* strain expressing Ste2-EGFP (JTY6757) and a *MATa* strain expressing Ste2-mCherry (YEL014) in the same manner were cultivated at either 20°C or 30°C. After incubation with fluorogen (0.4 mM dye; 15 min; pH 6.5), the cell populations were examined and compared by fluorescence microscopy. Representative images are shown for each strain and condition. Scale bar, 5 µm. (B) For the cell samples in A, PM-localized fluorescence was quantified (n > 250 cells each) using CellProfiler, and the values obtained were plotted in box-and-whisker format. Box represents the interquartile range (IQR) between lower quartile (25%) and upper quartile (75%); horizontal black line represents the median value; whisker ends represent the lowest and highest data points still within 1.5 IQR of the lower and upper quartiles, respectively; dot, a single cell that exhibited a fluorescence intensity higher than the upper quartile. For each strain, the initial median fluorescence intensity value at the PM obtained at 20°C was set to 100%. (C) The strains in A, as well as wild-type cells expressing Ste2-FLAG-(His)$_6$ (yAEA201) and an otherwise isogenic *yps1Δ mkc7Δ* strain expressing Ste2-FLAG-(His)$_6$ (yAEA361), were cultivated at either 20°C or 30°C, and extracts were prepared and samples (6 µg total protein) analyzed as in Figure 2B. (D) Left, the pheromone responsiveness of the indicated cultures from C was assessed using an agar diffusion (halo) bioassay to measure α-factor–induced growth arrest on BSM medium (15 µg α-factor spotted on each filter disk). Plates were incubated at the indicated temperature. Right, quantification of the average difference in halo diameter for the indicated strains (two biological and three technical replicates were performed for each) at 20° and 30°C. Error bars, SEM; **p value < 0.0001, determined by two-tailed Student's t test.

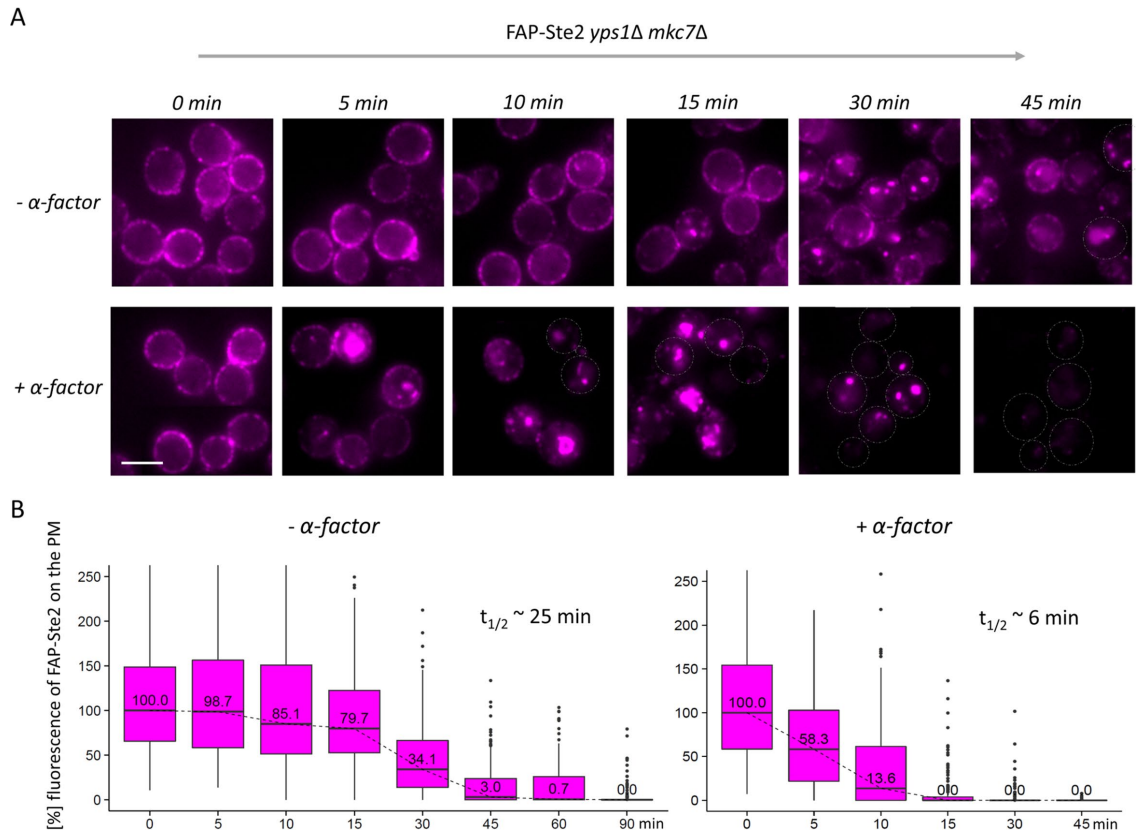

FIGURE 4: Direct visualization of basal and ligand-induced receptor internalization. (A) A MATa yps1Δ mkc7Δ strain expressing FAP-Ste2 (yAEA359) was grown at 20°C to early exponential phase, treated with LatA, incubated with fluorogen (0.4 mM dye; 15 min; pH 6.5), and deposited onto the glass bottoms of imaging chambers, and then internalization was initiated by washing out the LatA and excess fluorogen, as described under Materials and Methods, followed by either immediate addition of α-factor in H$_2$O (5 μM final concentration) (+α-factor) or an equivalent of water (–α-factor), and the cells were monitored by fluorescence microscopy at the indicated times over the course of 45–90 min. A representative image is shown for each time point. Scale bar, 5 μm. (B) The fluorescence intensity at the cell periphery in cells from the images (n = 5–6 per time point) from A were quantified using CellProfiler and plotted in box-and-whisker format, as in Figure 3B. For each strain, the initial median fluorescence intensity value at the PM was set to 100%. Insets, calculated times ($t_{1/2}$) for 50% decrease in PM fluorescence.

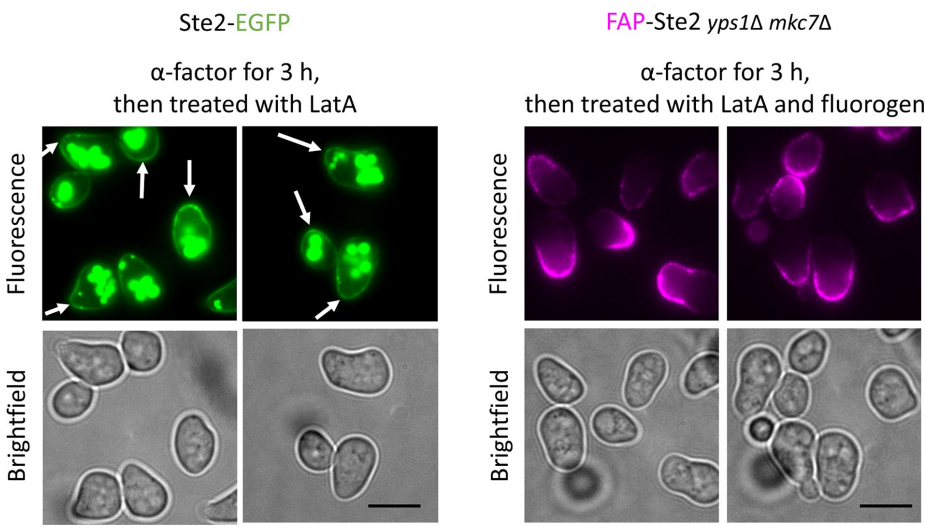

FIGURE 5: Cells expressing FAP-Ste2 exhibit a normal morphological response to α-factor and insert newly made receptors at the shmoo tip. MATa cells expressing Ste2-EGFP (JTY6765) (left) and MATa yps1Δ mkc7Δ cells expressing FAP-Ste2 (yAEA359) (right) were treated with 10 μM α-factor for 3 h, incubated with LatA (and, in case of FAP-Ste2, then with fluorogen), and examined by fluorescence microscopy. Scale bar, 5 μm. Arrows, very slight enrichment of Ste2-GFP at shmoo tips (as compared with the prominent FAP-Ste2 fluorescence at shmoo tips).

To address receptor fate in Glo3-deficient cells directly, we examined FAP-Ste2 and the dynamics of its pheromone-induced trafficking in MATa yps1Δ mck7Δ cells that either retained a functional GLO3 gene or carried a glo3Δ mutation and in which the rim of the vacuole was demarcated using Vph1-EGFP (Oku et al., 2017) (an integral membrane subunit of the V$_0$ component of the vacuolar ATPase), which we expressed under control of the VPH1$_{prom}$ but integrated at the HIS3 locus on chromosome XV. For the GLO3$^+$ cells, as we observed before (Figure 4), virtually no cells in the population had any FAP-Ste2 remaining at the PM by 45 min after exposure to α-factor and, as early as 15 min after addition of pheromone, readily detectable FAP fluorescence was observed within the lumen of the vacuole in every cell (Figure 6A, top panels). In marked contrast, in the *glo3Δ* cells, FAP-Ste2 persisted at the PM in a readily detectable fraction of the cells even 45 min after exposure to α-factor and, throughout the time course, very few of the cells contained detectable FAP fluorescence within the lumen of the vacuole (Figure 6A, bottom panels). Most strikingly, and as quantified in Figure 6B, the bulk of the FAP fluorescence in *glo3Δ* cells was confined to endosomes, often docked at or near the vacuole rim. Thus, unlike Kawada et al. (2015), we did not observe any drastic decrease in receptor level in cells lacking Glo3 (FAP-Ste2 at the PM in the *glo3Δ* mutant was at least 85% of that in isogenic *GLO3*+ cells), there was a noticeable decrease in the rate of receptor internalization in *glo3Δ* cells, and, most significantly, the primary defect in Ste2 trafficking in cells lacking Glo3 was in delivery of endosomes to the vacuole.

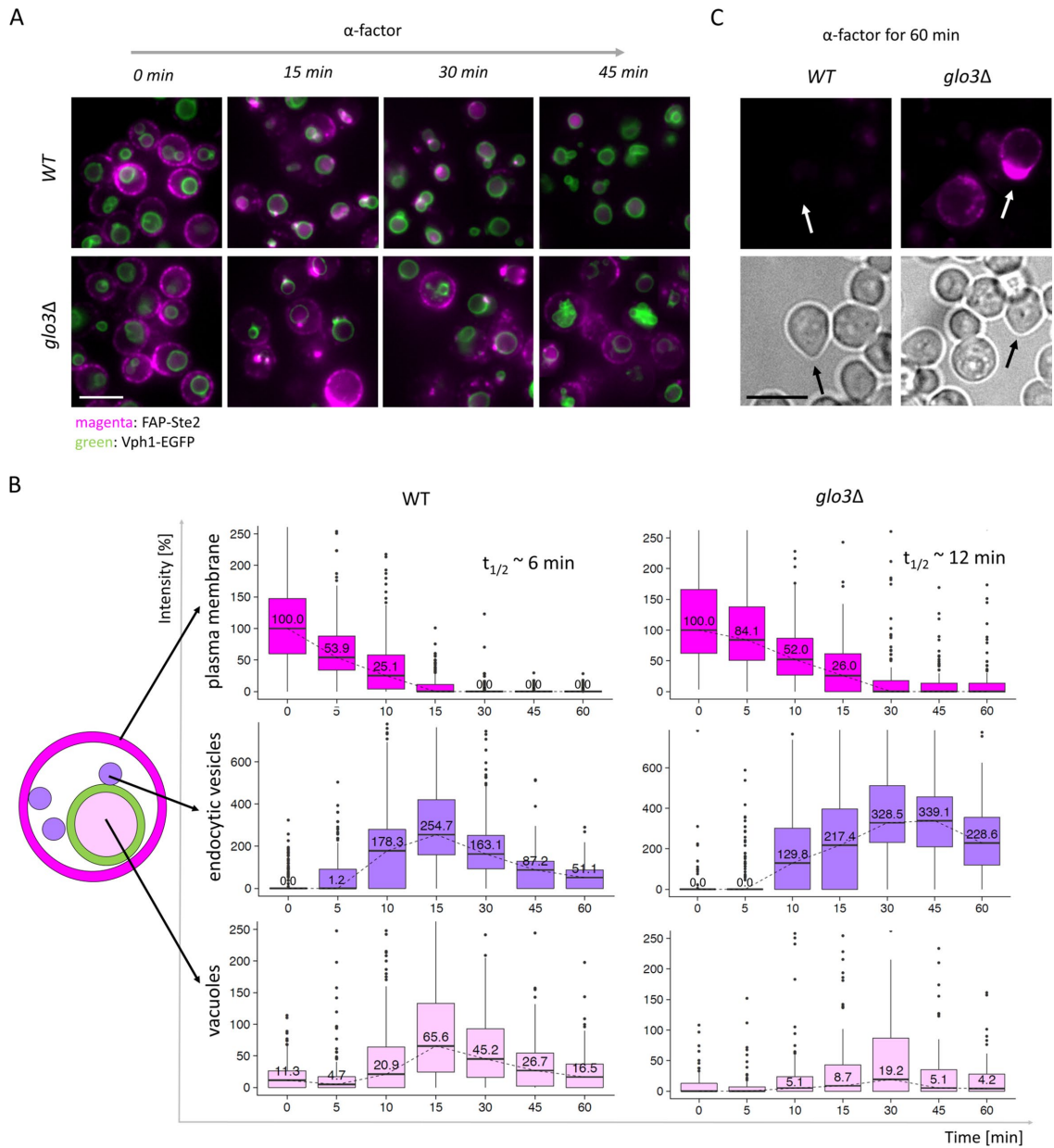

**FIGURE 6:** Delivery of Ste2 to the vacuole is defective in cells lacking Glo3. (A) Pheromone-induced endocytosis of FAP-Ste2 expressed in isogenic *GLO3*+(yAEA380) (top panels) and *glo3Δ* (yAEA382) (bottom panels) *MAT*a *yps1Δ mkc7Δ* Vph1-EGFP cells was conducted as in Figure 4. A representative image is shown for each strain at each time point. Scale bar, 5 μm. (B) The fluorescence intensity at the cell periphery (magenta), in endocytic vesicles (purple), and in the lumen of the vacuole (pink), as indicated in the schematic cell illustration to the left, in cells ($n \geq 250$) from the images ($n$ = 5–6 per time point) from A were quantified using CellProfiler and plotted in box-and-whisker format, as in Figure 3B. Insets, calculated times ($t_{1/2}$) for 50% decrease in PM fluorescence. For each strain, the initial median fluorescence intensity value at the PM was set to 100%. (C) The strains in A were grown to early exponential phase at 20°C, incubated with LatA and fluorogen, as described under *Materials and Methods*, washed, incubated with 10 μM α-factor in liquid medium for 60 min, and examined by fluorescence microscopy. Arrows, cells that have commenced forming shmoo tips. Scale bar, 5 μm.

As an independent means to document the delayed receptor internalization in the absence of Glo3; otherwise, wild-type

MAT**a** *yps1Δ mck7Δ* cells expressing FAP-Ste2 or the *glo3Δ* derivative were labeled with fluorogen and then exposed to excess α-factor for 1 h. Unlike the wild-type cells, a readily detectable portion of the population of *glo3Δ* mutant cells exhibited persistent FAP fluorescence at the cell surface (Figure 6C).

### The α-arrestins Ldb19, Rod1, and Rog3 play distinct roles in FAP-Ste2 internalization and post–endocytic sorting

Prior work has established that, of the 14 recognized *S. cerevisiae* α-arrestins, three (Ldb19 and apparent paralogues Rod1 and Rog3) contribute to down-regulation of Ste2 (Alvaro *et al.*, 2014, 2016; Prosser *et al.*, 2015). All three bind the E3 Rsp5, and Ste2 down-regulation by Ldb19 and Rod1 requires their interaction with Rsp5, whereas negative regulation of Ste2 by Rog3 does not obligatorily require its association with Rsp5 (Alvaro *et al.*, 2014). For MAT**a** *yps1Δ mkc7Δ* cells expressing FAP-Ste2, we found that loss of Ldb19 or of both Rod1 and Rog3, caused a modest, but reproducible, enhancement of their sensitivity to pheromone-induced growth arrest, as judged by the halo bioassay (Supplemental Figure S5), and the effect was maximal in the *ldb19Δ rod1Δ rog3Δ* triple mutant (hereafter "*3arrΔ*"), exactly as seen before for MAT**a** cells expressing wild-type Ste2 (Alvaro *et al.*, 2014). As we noted previously, given their different requirements, and because the effects of the absence of the three α-arrestins appear additive, this suggests that their contributions to receptor down-regulation may be exerted by different mechanisms.

To gain greater insight about how each of these α-arrestins contributes to the control of Ste2, we took two approaches. First, to assess the impact of the loss of all three α-arrestins on receptor behavior, we examined ligand-induced FAP-Ste2 internalization in MAT**a** *yps1Δ mkc7Δ* cells and otherwise isogenic MAT**a** *yps1Δ mkc7Δ 3arrΔ* cells (Figure 7A). We found that, in the absence of these three primary α-arrestins, α-factor-induced removal of FAP-Ste2 from the PM was not blocked, but its rate of internalization was slowed down by 50%, with a concomitant reduction in the rate with which FAP fluorescence appeared in endosomes (Figure 7B). It has been amply demonstrated that ubiquitinylation of seven Lys residues in the C-terminal cytosolic tail of Ste2 are mandatory for its endocytosis (Hicke and Riezman, 1996; Ballon *et al.*, 2006; Alvaro *et al.*, 2016). Likewise, we found that these same seven Lys residues were obligatory for FAP-Ste2 endocytosis (Figure 7C). Therefore, in the absence of Ldb19, Rod1, and Rog3, one or more of the remaining 11 α-arrestins, must be able, albeit less efficiently, to support Rsp5-mediated ubiquitinylation of FAP-Ste2 (and, normally, Ste2 itself). However, the most striking effect seen in the *3arrΔ* cells was a prolonged delay in the fusion of the endosomes, once formed, with the vacuole (Figure 7B); even at late times (e.g., 45 min after α-factor addition), the majority of the *3arrΔ* cells still had multiple endosomes docked at the vacuolar membrane, whereas very few of the control cells exhibited that pattern and had, by that time, degraded all the receptor (Figure 7A).

To complement the first approach and interrogate their individual roles in pheromone-induced endocytosis, each of the three α-arrestins (expressed from its native promoter on a *CEN* plasmid) (Table 1) was reintroduced into the *3arrΔ* cells. Because these proteins were untagged, we first examined their phenotypic effect on the pheromone sensitivity of the FAP-Ste2-expressing MAT**a** *yps1Δ mkc7Δ 3arrΔ* cells as a means to ensure that each was produced and functional. Reassuringly, expression of each α-arrestin, presumably at a near-endogenous level from the corresponding *CEN* plasmid, either partially reduced pheromone sensitivity (Rod1 and Rog3) or restored it to the level seen in wild-type control cells (Ldb19)

(Figure 8A). Therefore, the dynamics of FAP-Ste2 were examined after exposing the same three α-arrestin-expressing derivatives to α-factor (Figure 8B). Revealingly, restoration of Ldb19 alone markedly accelerated the rate of FAP-Ste2 endocytosis (reducing the $t_{1/2}$ for internalization from the PM from ~9 min down to ~4 min) and concomitantly increased the rate with which FAP-Ste2 appeared in endosomes and in the vacuole. The same trends were observed for the *3arrΔ* cells in which Rod1 was reintroduced (Figure 8B), but its effects were somewhat less pronounced than for Ldb19. Even though produced from their native promoters on a *CEN* plasmid, it is possible that the enhancement in the rate of FAP-Ste2 internalization observed in the *rod1Δ rog3Δ ldb19Δ* cells expressing either Rod1 or Ldb19 could arise from an elevation of the level of these proteins compared with that in WT cells. Nevertheless, these observations provide confirmation of prior, less-direct evidence (Alvaro *et al.*, 2014, 2016; Prosser *et al.*, 2015) that both Ldb19 and Rod1 act, at least in large measure, by promoting the earliest steps of cargo recognition and internalization by mediating efficient Rsp5-dependent ubiquitinylation of Ste2 at the PM.

In striking contrast, reintroduction of Rog3 markedly impeded the rate of α-factor-induced internalization of FAP-Ste2 and caused a pronounced delay in its appearance in endosomes and the vacuole. We have demonstrated using in vitro pull-down assays that Rog3 is able to bind to the cytosolic tail of Ste2 (Alvaro *et al.*, 2014). Thus, even though it associates with Rsp5, Rog3 itself must be unable to support sufficiently robust receptor ubiquitinylation to overcome the effect of the counteracting deubiquitinylating enzyme (Ubp2) (Kee *et al.*, 2005; Ho *et al.*, 2017), and, in addition, the presence of Rog3 must be able to block to a substantial degree whichever of the remaining 11 α-arrestins is responsible for the residual internalization observed in the *3arrΔ* cells. Indeed, even at very late times after pheromone addition (e.g., 45 min), and unlike *3arrΔ* cells expressing either Ldb19 alone or Rod1 alone, in many of the *3arrΔ* cells expressing Rog3 alone there persist endosomes that have not yet been fully delivered to the vacuole (Figure 8C), consistent with very slow or inefficient initial ubiquitinylation of the FAP-Ste2 cargo and/or an inability to maintain its ubiquitinylated state once internalized.

### DISCUSSION

Yeast has served as an invaluable model for dissecting the gene products and physiological processes that control the trafficking of proteins to (Schekman, 1995; Feyder *et al.*, 2015) and from (Goode *et al.*, 2015; Lu *et al.*, 2016) the PM. In this study, we were able to develop a tool to visualize, exclusively and for the first time, endocytic internalization of the preexisting surface-exposed pool of the endogenous GPCR Ste2 in yeast cells. A sensitive method is required because available estimates indicate that there are no more than 500 molecules of Ste2 per MAT**a** cell (Kulak *et al.*, 2014; Chong *et al.*, 2015). To do so required substantial refinement of the exocellular labeling method that utilizes the FAPα2 tag (Szent-Gyorgyi *et al.*, 2008; Fisher *et al.*, 2010). We found that a composite secretory signal (yeast MFα1$_{[1-83]}$-human Igκ signal peptide) worked best to maximize the amount of the FAP-receptor chimera at the PM, while preserving proper folding of both the FAP tag (as judged by the fluorescence intensity achieved on fluorogen binding) and receptor functionality (as judged by retention of responsiveness to the agonist, α-factor). Significantly, we found that stability of FAP-containing constructs in yeast required elimination of two, periplasmic, GPI-anchored aspartyl proteases, Yps1 and its paralogue Mkc7. In this same regard, in a report that just appeared describing the use of a FAP tag to track a mammalian potassium channel (Kir2.1) heterologously expressed in yeast cells, there is clear evidence based on the SDS–PAGE analysis shown that their FAP-Kir2.1 construct suffered proteolytic cleavage (Hager *et al.*, 2018). Given the number of transmembrane and extracellular proteases in mammalian cells (Overall and Blobel, 2007; Clark, 2014), our findings in yeast raise a note of caution about drawing conclusions using this approach in other organisms without first documenting that the initially produced FAP-tagged protein remains fully intact in the conditions under study.

Although removal of Yps1 and Mkc7 was required to maintain full-length FAP-Ste2, the absence of these two proteases did not have any deleterious effects on growth rate, cell morphology, or the behavior of FAP-Ste2 compared with Ste2 itself under our conditions. Nonetheless, absence of Yps1 and Mkc7 causes some changes in yeast cell wall composition (Krysan *et al.*, 2005). Our observations suggest these changes affect cell wall architecture and porosity. In otherwise wild-type *MAT***a** cells expressing either FAP-Ste2 or Ste2, an equivalent response was elicited by a given dose of pheromone, whereas for a *MAT***a** *yps1Δ mkc7Δ sst2Δ* strain the dose required to elicit an equivalent response from cells expressing FAP-Ste2 was approximately fourfold higher than for cells expressing Ste2. Similarly, although otherwise wild-type cells expressing FAP-Ste2 were able to bind A488-αF, for *MAT***a** *yps1Δ mkc7Δ* cells expressing FAP-Ste2 we were unable to detect any decoration with A488-αF (unpublished data), suggesting that the combination of the rather bulky fluorophore in A488-αF and the alteration of the cell wall caused by the absence of the two yapsins prevent diffusion of the fluorescent dye-tagged pheromone through the cell wall.

Likewise, unlike the rapid fluorogen labeling of the FAP tag on the surface of animal cells even on ice, we found that at least 15 min of incubation with fluorogen at an elevated temperature (30°C) and with some agitation were all required for optimal labeling of FAP-Ste2 expressed in *MAT***a** *yps1Δ mkc7Δ* cells, most likely to allow sufficient time for the dye to diffuse through the cell wall. Also, we found that growing the cells at 20°C and buffering the growth medium at pH 6.5 were critical for maximally efficient surface expression, fluorogen labeling, and retention of the fluorescent signal. When yeast cells grow on glucose in an unbuffered synthetic medium or in unbuffered rich yeast extract–peptone–dextrose (YPD) medium, the pH of the medium can drop to as low as 3.0–3.5 (Fraenkel, 2011), a condition under which it seems the FAP tag unfolds or misfolds. However, our experiments demonstrate that, once bound to fluorogen at pH 6.5, the FAP fluorescence remains stable within both endosomes and the vacuole, which are only mildly acidic compartments (Kane, 2006). The pH inside the yeast vacuole has been estimated to be between 6.2 (Preston *et al.*, 1989) and 5.3 (Brett *et al.*, 2011), values at which we still observed stable fluorogen binding. Thus, the eventual loss of the fluorescent signal inside the vacuole likely results from degradation of both the tag and the receptor portions of the FAP-Ste2 chimera by the vacuolar proteases, in agreement with prior work demonstrating that destruction of both Ste2 and its bound ligand are blocked in mutants lacking Pep4/Pra1 (Schimmöller and Riezman, 1993; Schandel and Jenness, 1994), a vacuolar proteinase required to mature the precursors to the other major vacuolar proteases (Jones, 2002). In any event, being alert to each of the concerns summarized above allowed us to productively utilize the FAP technology to examine a variety of aspects of Ste2 dynamics that had heretofore been inaccessible to experimental interrogation. Indeed, FAP-Ste2 always yielded much brighter and distinct fluorescent signals, allowing for better visualization and quantification, compared with Ste2-EGFP or Ste2-mCherry, which are plagued by massive background fluorescence accumulated in the vacuole (Suchkov *et al.*, 2010; Venkatapurapu *et al.*, 2015).

Using our FAP-Ste2 probe, we ascertained that the absence of the Arf-GAP Glo3 affects receptor trafficking in ways different from those initially deduced from monitoring the behavior of a fluorescent α-factor derivative or radioactive α-factor as proxies for the receptor (Kawada *et al.*, 2015). Other work (Poon *et al.*, 1999; Bao *et al.*, 2018) has established that Glo3 is involved in controlling retrograde transport from the Golgi compartment back to the endoplasmic reticulum. Kawada *et al.* (2015) observed that, in cells lacking Glo3, there was a drastic reduction in pheromone binding at the cell surface with a concomitant increase in the amount of pheromone in the vacuole, suggesting that the Ste2 can be internalized but not efficiently recycled to the PM. However, using FAP-Ste2 to visualize the receptor itself, we did not find any drastic decrease in receptor level at the PM in cells lacking Glo3, and the major defect was prolonged delay in the delivery of FAP-Ste2-containing endosomes to the vacuole. Moreover, although the rate of basal endocytosis of FAP-Ste2 is much slower than the rate of its pheromone-induced internalization (as observed for native Ste2), under either condition, all of the endocytosed FAP-Ste2 is eventually delivered to the vacuole with no detectable recycling to the PM.

As another test of the utility of this approach, we used our FAP-Ste2-expressing *MAT***a** *yps1Δ mkc7Δ* cells to address the individual roles of three endocytic adaptors, the α-arrestins Ldb19/Art1, Rod1/Art4, and Rog3/Art7, that we had previously shown are involved in down-regulation of Ste2-initiated signaling (Alvaro *et al.*, 2014, 2016; Prosser *et al.*, 2015). As observed before for wild-type cells expressing native Ste2, we found modest but readily detectable and reproducible increases in pheromone sensitivity (as judged by the diameter of the halo of G1-arrested cells) for *MAT***a** *yps1Δ mkc7Δ* expressing FAP-Ste2 that lacked Ldb19 or both Rod1 and Rog3, or all three (*3arrΔ* mutant), despite the fact that these cells possess all of the previously characterized mechanisms for recovery and adaptation that act at the receptor level, described in the Introduction, as well as those that act at more distal points in the pheromone response signaling pathway (Dohlman and Thorner, 2001; Alvaro and Thorner, 2016). To better understand how each of these three α-arrestins contributes to down-regulation of pheromone signaling, we reintroduced each of them into *MAT***a** *yps1Δ mkc7Δ 3arrΔ* cells expressing FAP-Ste2. Strikingly, we found that presence of Ldb19 alone or Rod1 alone accelerated initial pheromone-induced internalization to a rate that was approximately twofold faster than that observed even in wild-type cells, suggesting that each of these α-arrestins works better to mediate Rsp5-dependent ubiquitinylation of the receptor in the absence of competition from the other two. Even more revealingly, despite the fact that Rod1 and Rog3 share greater similarity to each other (45% identity) than to any other *S. cerevisiae* α-arrestin, reintroduction of Rog3 alone markedly impeded the rate of pheromone-induced internalization. The latter finding is consistent with and greatly extends prior, less direct evidence (Alvaro *et al.*, 2014, 2016) that Rog3-imposed inhibition of receptor signaling does not require its association with Rsp5 and that Rog3 is an "Ur" β-arrestin-like regulator, namely blocking signaling by occluding receptor association with its cognate heterotrimeric G-protein, rather than stimulating receptor ubiquitinylation and internalization *per se*. Alternatively, because there are reports that Ste2 is internalized as at least a dimer or higher oligomer (Overton and Blumer, 2000; Yesilaltay and Jenness, 2000), Rog3 binding may prevent the receptor self-association necessary to form dimers or higher-order complexes.

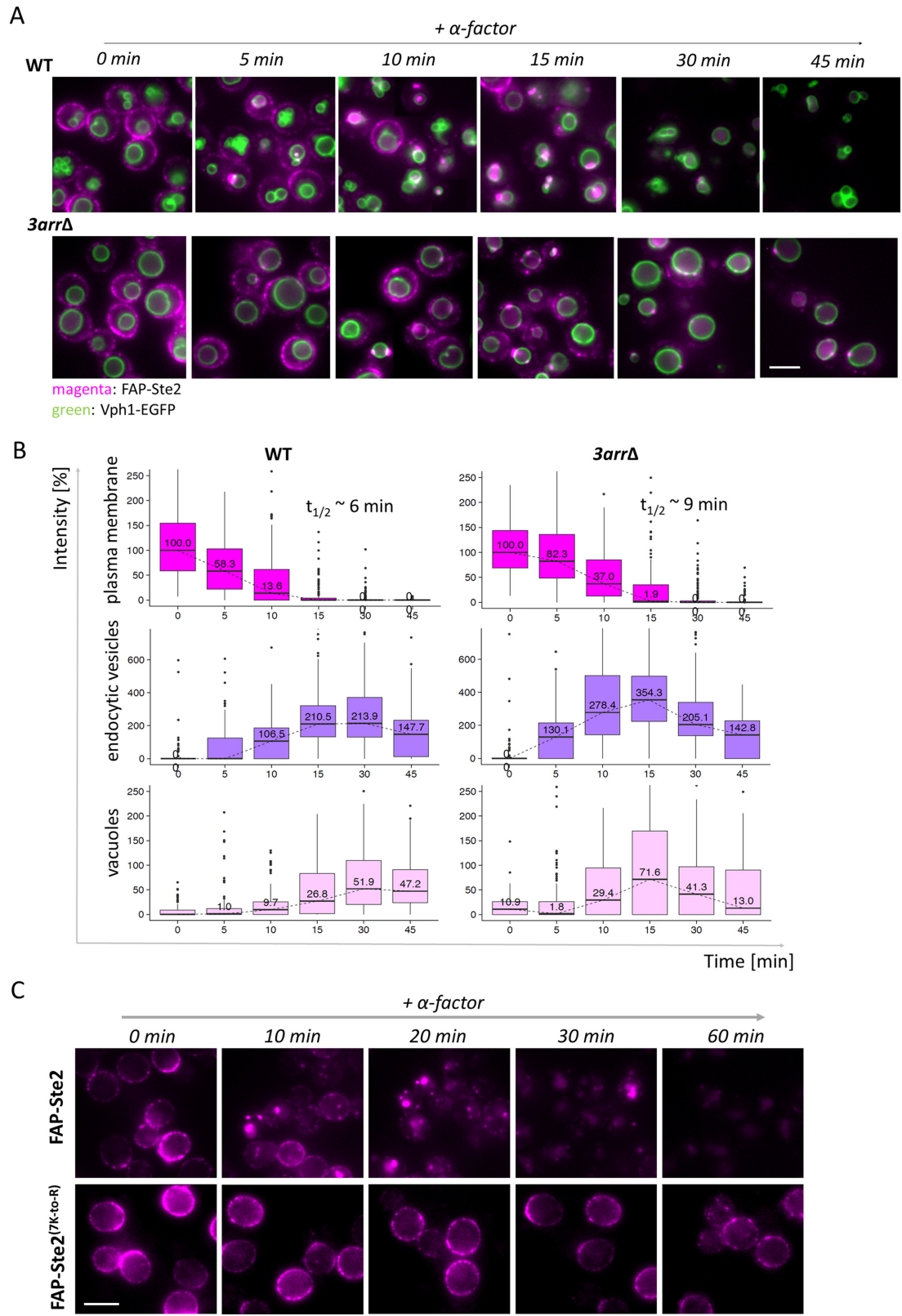

FIGURE 7: Absence of α-arrestins Ldb19, Rod1, and Rog3 delays internalization and delivery of endocytosed FAP-Ste2 to the vacuole. (A) Otherwise isogenic *MATa* (yAEA380) and *MATa 3arrΔ* (yAEA381) cells expressing FAP-Ste2 and Vph1-EGFP were cultivated and incubated with 5 µM α-factor to initiate pheromone-induced endocytosis as described in Figure 6A. A representative image is shown for each strain at each time point. Scale bar, 5 µm. (B) The data in A were quantified and plotted as described in Figure 6B. The initial intensities of FAP-Ste2 on the PM (i.e., at time 0) were quite similar for both strains, and their median values were set to 100%. (C) *MATa yps1Δ mkc7Δ* cells expressing either FAP-Ste2 (yAEA359) or FAP-Ste2(7K-to-R) (yAEA397) were grown at 20 °C to early exponential phase, treated with LatA, incubated with fluorogen (0.4 mM dye; 15 min; pH 6.5), and deposited onto the glass bottoms of imaging

| Plasmid | Genotype | Reference or source |
|---|---|---|
| pRS316 | CEN URA3 | Sikorski and Hieter, 1989 |
| pJT4439 | pRS316-LDB19$_{prom}$-LDB19 CEN URA3 | C. Alvaro, this lab |
| pJT4436 | pRS316-ROG3$_{prom}$-ROG3 CEN URA3 | C. Alvaro, this lab |
| pJT4436 | pRS316-ROD1$_{prom}$-ROD1 CEN URA3 | C. Alvaro, this lab |
| pNH603 | Derivative of pRS303 | Sikorski and Hieter, 1989; Moser et al., 2013 |
| pUB691 | pNH603-HIS3::VPH1-EGFP | Gift of Yuzhang Chen and Elçin Ünal, UC Berkeley |

TABLE 1: Plasmids used in this study.

Unexpectedly, we found that when Ldb19 was absent there was a more pronounced accumulation of FAP-Ste2-containing endosomes, many of which appeared to be docked on the vacuole membrane. Ldb19 was first found to contribute to the efficient downregulation of several amino acid permeases (Mup1, Can1, and Lyp1) (Lin et al., 2008; Nikko and Pelham, 2009). To date, however, the current evidence is unclear about the exact subcellular location of this α-arrestin. Ldb19/Art1 C-terminally tagged with GFP has been found diffusely in the cytosol but also in punctate structures that may or may not be the late Golgi compartment and also at the cell cortex associated with the plasma membrane and/or early endosomes (Huh et al., 2003; MacGurn et al., 2011). Our results using FAP-Ste2 raise the possibility that sustained Ldb19-dependent Rsp5-mediated ubiquitinylation on endosomes may be required to ensure efficient cargo recognition for ESCRT-mediated delivery of these endosomes to the MVB/vacuole. This conclusion is at least consistent with recent evidence that, for endosomes containing the lactate permease Jen1, Rod1 seems to be required mainly for their post–endocytic sorting to the vacuole rather than for the initial internalization of Jen1 (Becuwe and Léon, 2014; Hovsepian et al., 2018) and that other α-arrestins have roles in intracellular trafficking separate from their function in the initial steps of endocytosis (Risinger and Kaiser, 2008; O'Donnell et al., 2010; O'Donnell, 2012). Taken together, our findings indicate that different α-arrestins act differentially and at distinct stages along the endocytic pathway to control receptor signaling and homeostasis.

There are many additional questions about receptor dynamics that can now be addressed readily using FAP-Ste2. Moreover, we hope that our developing the insights and conditions needed to apply this method productively in yeast will allow other investigators to interrogate the behavior of integral PM proteins of greatest interest to them. However, the FAP tag is not a panacea for monitoring the dynamics of every integral PM protein. Our work revealed some limitations for its use in yeast. The need for the yps1Δ mkc7Δ double mutant background could complicate some experimental designs because such cells are temperature sensitive, grow poorly at low pH, and exhibit elevated sensitivity to a number of drugs and other stressful conditions (Komano and Fuller, 1995; Krysan et al., 2005; Cho et al., 2010). These phenotypes might preclude use of the FAP tag for analysis of some endocytic cargos or in some mutants that affect the endocytic pathway. Also, for polytopic PM proteins in which both the N and C termini face the cytosol, the FAP tag would need to be inserted into an extracellular loop, which might interfere with folding or function of either the protein and/or the tag.

## MATERIALS AND METHODS
### Cloning and strain construction
Constructs used for cassette amplification were assembled using standard procedures (Green and Sambrook, 2012). DNAs encoding the FAPα1 and FAPβ2 tags were purchased from SpectraGenetics (Pittsburgh, PA) and fused in-frame to the initiator ATG at the N terminus of the STE2 ORF, which was tagged at its C terminus with a FLAG epitope and (His)$_6$ tract (David et al., 1997), as described in detail in the Supplemental Material. PCR amplification was performed using Phusion DNA polymerase (New England BioLabs, Ipswich, MA), and all constructs were verified by DNA sequencing. Standard genetic methods were used for strain construction (Amberg et al., 2005). Correct integration of expression cassettes into the yeast genome were confirmed by colony PCR and sequencing.

### Growth conditions and incubation with fluorogen
Yeast strains (Table 2) were grown at 20°C (unless otherwise indicated) in a buffered synthetic media (BSM) (2% glucose, 5 mg/ml casamino acids, 1.7 mg/ml yeast nitrogen [without either ammonium sulfate or amino acids], 5.3 mg/ml (NH$_4$)$_2$SO$_4$, 20 μg/ml uracil, 100 mM Na phosphate [pH 6.5]) to an A$_{600nm}$ = 0.5. For fluorogen binding, cells (0.75 A$_{600nm}$ equivalent) were collected by brief centrifugation (30 s at 5000 rpm) and resuspended in 20 μl of fresh BSM. When indicated, LatA (Cayman Chemical, Ann Arbor, MI) was added (100 μM final concentration), and after incubation for 5 min at 30°C, 5 μl of a 2 mM stock of fluorogen, the cell-impermeable malachite green derivative αRED-np (SpectraGenetics), was added. After incubation with agitation (1200 rpm) for 15 min at 30°C in a Thermomixer (Eppendorf AG, Hamburg, Germany), the cells were recollected by brief centrifugation, washed twice by resuspension and brief recentrifugation in 1 ml ice-cold BSM, resuspended in 20 μl of ice-cold BSM, and used immediately to initiate experiments (or kept on ice for no longer than 30 min before use).

### Immunoblot analysis
Cells from early-exponential-phase cultures (10 A$_{600nm}$ equivalent) were collected by centrifugation and lysed, and the total membrane fraction was isolated as described previously (David et al., 1997). Membrane pellets were dispersed by trituration in a micropipette with 60 μl of 50 mM Na-phosphate (pH 7.5), and protein chambers, and then internalization was initiated by washing out the LatA and excess fluorogen, as described under *Materials and Methods*, followed by immediate addition of α-factor in H$_2$O (5 μM final concentration), and the cells were monitored by fluorescence microscopy at the indicated times over the course of 60 min. A representative image is shown for each time point. Scale bar, 5 μm.

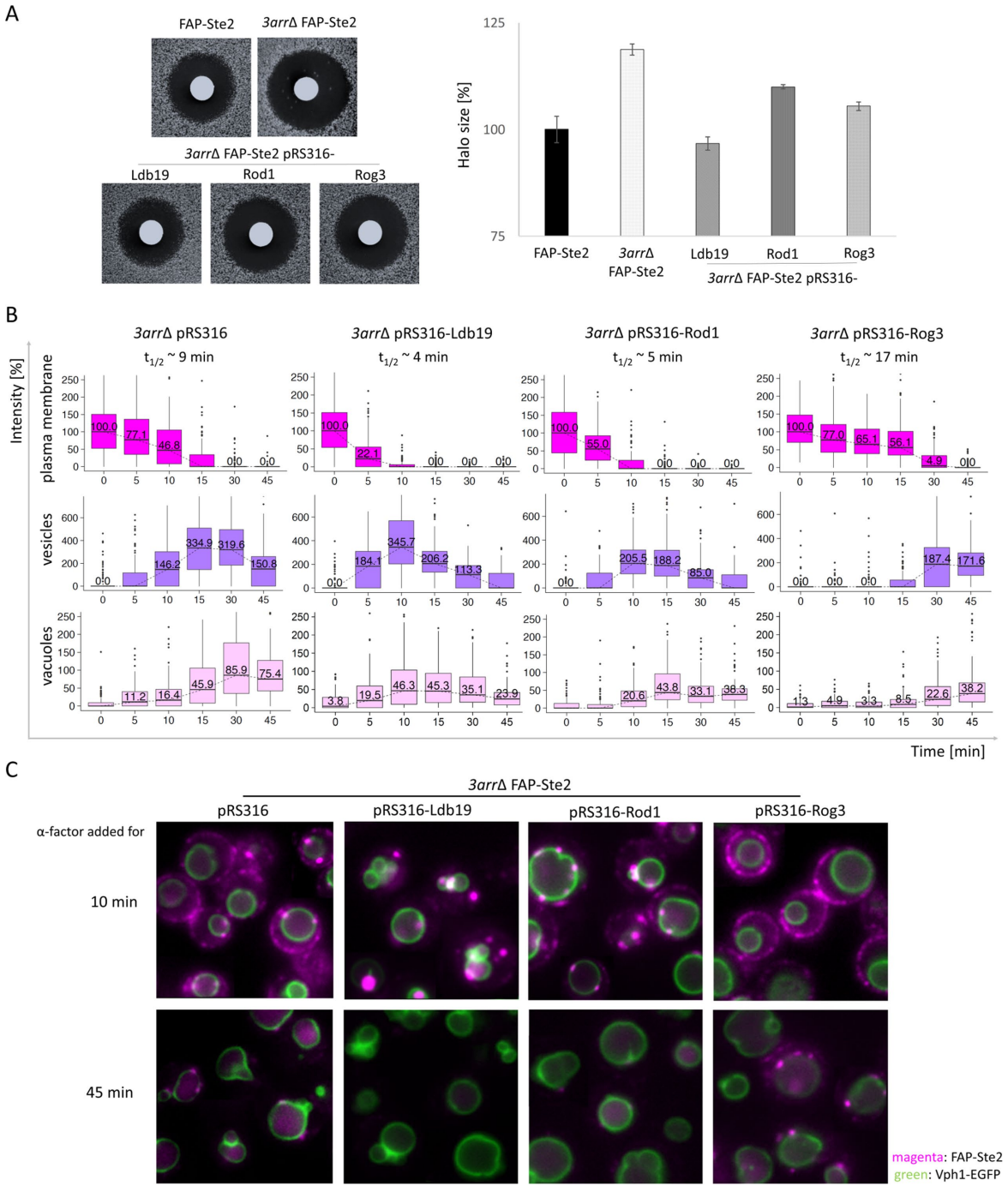

**FIGURE 8:** Ldb19, Rod1, and Rog3 have distinct roles in Ste2 down-regulation. (A) Left, the halo bioassay for pheromone-induced growth arrest was used to assess the relative pheromone sensitivity of a *MAT*a *yps1Δ mkc7Δ* FAP-Ste2 Vph1-EGFP strain (yAEA380) and an otherwise isogenic *3arrΔ* derivative (yAEA389), both carrying empty vector (pRS316) (top panels), as well as the same *3arrΔ* strain expressing *LDB19*, *ROD1*, or *ROG3*, as indicated, from the same vector (bottom panels), as in Figure 3D, except that the medium was BSM-Ura and 15 μg α-factor were spotted on the filter disks. Right, results of independent experiments ($n = 6$) are plotted as a bar graph, as in Figure 3D. (B) The same strains as in A were labeled with fluorogen, exposed to α-factor, and examined by fluorescence microscopy, as in Figure 3A, and the data were analyzed and plotted as in Figure 6B. The initial intensities of FAP-Ste2 on the PM (i.e., at time 0) were very similar for all four strains, and their median values were set to 100%; $t_{1/2}$, calculated time for 50% decrease in PM fluorescence. (C) Representative images for the strains in B at the indicated time points. Scale bar, 2.5 μm.

concentration was estimated using a commercial Bradford protein assay kit (BioRad, Hercules, CA). An appropriate volume of each resuspended pellet (6 μg total protein) was transferred to a fresh tube, collected by sedimentation at 15,000 rpm for 15 min at 4°C, and solubilized in 10 μl of 2 × SDS–urea sample buffer (6% SDS, 6 M urea, 25% glycerol, 5% 2-mercaptoethanol, a trace of bromophenol blue, 150 mM Tris-HCl [pH 6.8]). The solubilized proteins were resolved by SDS–PAGE and transferred electrophoretically to nitrocellulose membranes (Towbin et al., 1979) using a wet transfer apparatus (Bio-Rad). After blocking with Odyssey Blocking Buffer (in phosphate-buffered saline) (Li-Cor, Lincoln, NE) for 1 h, the membranes were incubated overnight at 4°C with an appropriate antibody: mouse anti-HA mAb 6E2 (Cell Signaling Technology, Danvers, MA), mouse anti-Pma1 mAb 40B7 (Abcam, Cambridge, MA), or rabbit polyclonal anti-Ste2 antibodies (raised against the C-terminal 131 residues of Ste2 (David et al., 1997). After washing with TBS-1% Tween, immune complexes on the membranes were detected by incubation with an appropriate infrared dye–(IRDye 680/800)-labeled secondary antibody, either goat-anti-mouse IgG or goat anti-rabbit IgG (Li-Cor), and scanned using an Odyssey CLx infrared imager (Li-Cor). Molecular weight markers used were the PageRuler prestained protein ladder (Crystalgen, Commack, NY).

| Strain | Genotype | Reference or source |
|---|---|---|
| BY4741 | MATa leu2Δ0 ura3Δ0 his3Δ1 met15Δ0 | Research Genetics |
| JTY4470 | BY4741 ste2Δ::KanMX4 | Research Genetics |
| yAEA201 | BY4741 STE2$_{prom}$::STE2-FLAG-(His)$_6$::URA3 | This study |
| yAEA152 | BY4741 STE2$_{prom}$::MFα1$_{(1-83)}$-Igκ-FAPα2-STE2-FLAG-(His)$_6$::URA3 [FAP-Ste2] | This study |
| yAEA265 | BY4741 STE2$_{prom}$::STE2-FLAG-(His)$_6$::URA3 bar1Δ::KanMX | This study |
| yAEA261 | BY4741 STE2$_{prom}$::MFα1$_{(1-83)}$-Igκ-FAPα2-STE2-FLAG-(His)$_6$::URA3 bar1Δ::KanMX4 | This study |
| DK102 | BY4741 ste2Δ::HIS3 bar1Δ | D. Kaim, this lab |
| yAEA361 | BY4741 STE2$_{prom}$::STE2-FLAG-(His)$_6$::URA3 yps1Δ::KanMX mkc7Δ::KanMX | This study |
| yAEA363 | BY4741 STE2$_{prom}$::MFα1$_{(1-83)}$-Igκ-FAPα2-STE2-FLAG-(His)$_6$::URA3 yps1Δ::KanMX | This study |
| yAEA365 | BY4741 STE2$_{prom}$::MFα1$_{(1-83)}$-Igκ-FAPα2-STE2-FLAG-(His)$_6$::URA3 mkc7Δ::KanMX | This study |
| yAEA359 | BY4741 STE2$_{prom}$::MFα1$_{(1-83)}$-Igκ-FAPα2-STE2-FLAG-(His)$_6$::URA3 yps1Δ::KanMX mkc7Δ::KanMX | This study |
| JTY6757 | BY4741 STE2$_{prom}$::STE2-EGFP::HphNT1 | Alvaro et al., 2014 |
| YEL014 | BY4741 STE2$_{prom}$::STE2-mCherry::CaURA3 | E. Sartorel, this lab |
| YDB103 | BY4741 sst2Δ::KanMX ste2Δ | Ballon et al., 2006 |
| yAEA260 | BY4741 STE2$_{prom}$::STE2-FLAG-(His)$_6$::URA3 sst2::HphNT1 | This study |
| yAEA372 | BY4741 STE2$_{prom}$::STE2-FLAG-(His)$_6$::URA3 yps1Δ::KanMX mkc7Δ::KanMX sst2::HphNT1 | This study |
| yAEA373 | BY4741 STE2$_{prom}$::MFα1$_{(1-83)}$-Igκ-FAPα2-STE2-FLAG-(His)$_6$::URA3 yps1Δ::KanMX mkc7Δ::KanMX sst2::HphNT1 | This study |
| yAEA257 | BY4741 STE2$_{prom}$::STE2-EGFP::HphNT1 sst2Δ::KanMX | This study |
| yAEA258 | BY4741 STE2$_{prom}$::STE2-mCherry::CaURA3 sst2Δ::HphNT1 | This study |
| yAEA382 | BY4741 STE2$_{prom}$::MFα1$_{(1-83)}$-Igκ-FAPα2-STE2-FLAG-(His)$_6$::URA3 yps1Δ::KanMX mkc7Δ::KanMX glo3Δ::HphNT1 VPH1-EGFP::HIS3 | This study |
| yAEA379 | BY4741 STE2$_{prom}$::MFα1$_{(1-83)}$-Igκ-FAPα2-STE2-FLAG-(His)$_6$::URA3 yps1Δ::KanMX mkc7Δ::KanMX ldb19Δ::NatMX rog3Δ::KanMX rod1Δ::KanMX | This study |
| yAEA380 | BY4741 STE2$_{prom}$::MFα1$_{(1-83)}$-Igκ-FAPα2-STE2-FLAG-(His)$_6$::URA3 yps1Δ::KanMX mkc7Δ::KanMX VPH1-EGFP::HIS3 | This study |
| yAEA381 | BY4741 STE2$_{prom}$::MFα1$_{(1-83)}$-Igκ-FAPα2-STE2-FLAG-(His)$_6$::URA3 yps1Δ::KanMX mkc7Δ::KanMX ldb19Δ::NatMX rog3Δ::KanMX rod1Δ::KanMX VPH1-EGFP::HIS3 | This study |
| yAEA383 | BY4741 STE2$_{prom}$::MFα1$_{(1-83)}$-Igκ-FAPα2-STE2-FLAG-(His)$_6$::URA3 yps1Δ::KanMX mkc7Δ::KanMX ldb19Δ::NatMX VPH1-EGFP::HIS3 | This study |
| yAEA384 | BY4741 STE2$_{prom}$::MFα1$_{(1-83)}$-Igκ-FAPα2-STE2-FLAG-(His)$_6$::URA3 yps1Δ::KanMX mkc7Δ::KanMX rod1Δ::KanMX VPH1-EGFP::HIS3 | This study |
| yAEA385 | BY4741 STE2$_{prom}$::MFα1$_{(1-83)}$-Igκ-FAPα2-STE2-FLAG-(His)$_6$::URA3 yps1Δ::KanMX mkc7Δ::KanMX rog3Δ::KanMX VPH1-EGFP::HIS3 | This study |
| yAEA388 | BY4741 STE2$_{prom}$::MFα1$_{(1-83)}$-Igκ-FAPα2-STE2-FLAG-(His)$_6$::URA3 yps1Δ::KanMX mkc7Δ::KanMX rod1Δ::KanMX rog3Δ::KanMX VPH1-EGFP::HIS3 | This study |
| yAEA389 | BY4741 STE2$_{prom}$::MFα1$_{(1-83)}$-Igκ-FAPα2-STE2-FLAG-(His)$_6$::HphNT1 yps1Δ::KanMX mkc7Δ::KanMX ldb19Δ::NatMX rog3Δ::KanMX rod1Δ::KanMX VPH1-EGFP::HIS3 | This study |
| yAEA397 | BY4741 STE2$_{prom}$::MFα1$_{(1-83)}$-Igκ-FAPα2-STE2(7K-to-R)-FLAG-(His)$_6$::URA3 yps1::KanMX mkc7::KanMX [FAP-Ste2(7K-to-R)] | This study |

TABLE 2: Yeast strains used in this study.

Response to α-factor was assessed using an agar diffusion (halo) bioassay (Reneke et al., 1988; Alvaro et al., 2014). In brief, MATa cells (~10$^5$) of the indicated genotype were plated in top-agar on solid BSM or BSM-Ura medium, as appropriate. On the resulting surface were laid sterile cellulose filter disks onto which an aliquot (typically 15 µl) of a 1-mg/ml solution of α-factor (GeneScript, Piscataway, NJ) had been aseptically spotted, and the plates were incubated at 30°C for 2 d. For dose–response curves, a range of α-factor concentrations (0.125–30 µg per disk) were used, and the MATa cells carried an sst2Δ mutation to enhance pheromone sensitivity (Chan and Otte, 1982; Dohlman et al., 1996).

### Receptor-mediated endocytosis of Alexa 488-α-factor

Alexa Fluor 488–labeled α-factor was generously provided by David G. Drubin (University of California [UC], Berkeley) and internalization studies were performed by minor modifications of the procedure previously described (Toshima et al., 2006). Briefly, MATa bar1Δ cells were grown to an $A_{600nm}$ of 0.3–0.5 at 20°C in BSM and a sample (0.75 $A_{600nm}$ equivalent) was collected by brief centrifugation (30 s at 5000 rpm), washed once by resuspension in 1 ml ice-cold glucose-free BSM with 1% (wt/vol) bovine serum albumin (BSA), recollected by centrifugation, and resuspended in 20 µl glucose-free BSM with 1% (wt/vol) BSA, and A488-αF (5 µM final concentration) was added. After incubation on ice for 1.5 h, cells were washed three times with 1 ml ice-cold glucose-free BSM with 1% (wt/vol) BSA, resuspended in 500 µl of BSM containing 2% glucose, incubated at 30°C for indicated times, then fixed by addition of 10% (vol/vol) of 37% formaldehyde, and, after incubation for 1 h at room temperature, examined by fluorescence microscopy.

### Live-cell imaging of FAP-Ste2 internalization and image analysis

MATa cells of the indicated genotype expressing FAP-Ste2 were grown at 20°C to an $A_{600nm}$ = 0.3–0.5 at 20°C in BSM, treated with 100 µM LatA, and incubated with 0.4 mM fluorogen, as described above, then deposited onto the surface of the glass bottom of a 35-mm-well imaging dish (Integrated BioDiagnostics [ibidi] GmbH, Martinsried, Germany) that had been precoated with concanavalin A (0.1 µg/ml). After the well was rinsed three times with 1 ml BSM at room temperature, cells were overlaid with 1 ml BSM and incubated at 30°C for 20 min to allow for recovery from the LatA treatment. For pheromone-induced endocytosis, synthetic α-factor (GeneScript, Piscataway, NJ) was then added (usually 5 µM final concentration, unless otherwise indicated), and the cells were incubated at room temperature and examined by fluorescence microscopy at various times thereafter. Fluorescence microscopy was performed using an Elyra PS.1 structured illumination (SIM) microscope (Carl Zeiss AG, Jena, Germany) equipped with a 100× PlanApo 1.46NA TIRF objective, a main focus drive of the AxioObserver Z1 Stand, a WSB PiezoDrive 08, controlled by Zen, and images were recorded using a 512 × 512 (100 nm × 100 nm pixel size) electron-multiplying charge-coupled device (EM-CCD) camera (Andor Technology, South Windsor, CT). To visualize FAP-Ste2 (excitation $\lambda_{max}$ 631 nm; emission $\lambda_{max}$ 650), samples were excited with an argon laser at 642 nm at 2.3% power (100 mW), and emission was filtered at >655 nm; for EGFP-tagged proteins (excitation $\lambda_{max}$ 489 nm; emission $\lambda_{max}$ 508), excitation was at 488 nm at 2.3% power (100 mW) and emission monitored in a 495–550 nm window using a bandpass filter; for Ste2-mCherry (excitation $\lambda_{max}$ 587 nm; emission $\lambda_{max}$ 610), excitation was at 561 nm at 2.3% power (100 mW), and emissions were monitored in a 570- to 620-nm window using a different bandpass filter. Images (average of eight scans; 300 ms/scan) were analyzed using Fiji (Schindelin et al., 2012). To avoid changes in image quality due to occasional fluctuations in laser intensity, all panels shown in any given figure represent experiments performed on the same day and are scaled and adjusted identically for brightness using Fiji (Schindelin et al., 2012). For quantitative automated analysis of fluorescence intensity at the PM, in endosomes, or in the vacuole lumen, CellProfiler was used (Carpenter et al., 2006). To train CellProfiler to apply appropriate masks and separately quantify the signal from each of these compartments, a corresponding pipeline was created, which was adapted from prior software (Bray et al., 2015; Chong et al., 2015) (Supplemental File 1). Prior to loading into the CellProfiler pipeline, cell images were segmented manually using Fiji (Schindelin et al., 2012). To avoid any selection bias, every cell visible in the bright field image in a frame from any sample (except those out-of-focus) was chosen. All plots and statistical analyses in this study were performed with R (R Core Team, 2018).

### Reproducibility

All results reported reflect, except where indicated otherwise, findings repeatedly made in at least three independent trials of each experiment shown. Sample sizes, number of biological and technical replicates performed, statistical analysis used, and whether and how the values presented were normalized are all described in the relevant figure legends.

## ACKNOWLEDGMENTS

This work was supported by Erwin Schroedinger Fellowship J3787-B21 from the Austrian Science Fund (to A.E.-A.), by a Marie Sklodowska-Curie Individual Fellowship GAND35 from the European Commission (to C.M.A.), and by National Institutes of Health (NIH) R01 Research Grant GM21841 (to J.T.). This work was also aided, in part, by NIH S10 Equipment Grant OD018136 (to Steven E. Ruzin, Director, UC Berkeley Biological Imaging Facility) for a Zeiss Elyra S1 structured illumination microscope. We thank Allyson F. O'Donnell (University of Pittsburgh) and especially Jonathan W. Jarvik (Carnegie Mellon University) for their insights about FAP technology, Robert S. Fuller (University of Michigan, Ann Arbor) for his expertise with yapsins, Steve Ruzin and Denise Schichnes (UC Berkeley) for their invaluable advice about fluorescence microscopy, and all members of the Thorner Lab (but especially Françoise Roelants) for helpful discussions. We are grateful to Ross T. A. Pedersen (Drubin lab, UC Berkeley) for assistance with the use of Alexa Fluor 488–labeled α-factor. This article is dedicated to the memory of the late Christopher G. ("Cris") Alvaro, a gifted scientist and exemplary human being.